\documentclass[prl, lettersize,twocolumn,superscriptaddress,
preprintnumbers,nofootinbib,showpacs]{revtex4-1}
\usepackage{graphicx}
\usepackage{amsmath}
\usepackage{amssymb}
\usepackage{color}
\usepackage{hyperref}
\usepackage{url}
\usepackage{breakurl}
\usepackage{float}
\usepackage{epstopdf}
\usepackage{extarrows}
\usepackage[normalem]{ulem}
\usepackage{multirow}

\newcommand{\beq}{\begin{equation}}
\newcommand{\eeq}{\end{equation}}
\newcommand{\be}{\begin{equation}}
\newcommand{\ee}{\end{equation}}
\newcommand{\bea}{\begin{eqnarray}}
\newcommand{\eea}{\end{eqnarray}}
\newcommand{\nn}{\nonumber}

\newcommand{\ld}{\lambda}

\allowdisplaybreaks[4]

\begin{document}
\preprint{
{\vbox {
\hbox{\bf LA-UR-20-23742}
\hbox{\bf MSUHEP-20-004}
}}}
\vspace*{0.2cm}

\title{Probing $Zt\bar{t}$ couplings using $Z$ boson polarization in $ZZ$ production at hadron colliders}

\author{Qing-Hong Cao}
\email{qinghongcao@pku.edu.cn}
\affiliation{Department of Physics and State Key Laboratory of Nuclear Physics and Technology, Peking University, Beijing 100871, China}
\affiliation{Collaborative Innovation Center of Quantum Matter, Beijing 100871, China}
\affiliation{Center for High Energy Physics, Peking University, Beijing 100871, China}

\author{Bin Yan}
\email{binyan@lanl.gov}
\affiliation{Theoretical Division, Group T-2, MS B283, Los Alamos National Laboratory, P.O. Box 1663, Los Alamos, NM 87545, USA}

\author{C.-P. Yuan}
\email{yuanch@msu.edu}
\affiliation{Department of Physics and Astronomy, Michigan State University, East Lansing, MI 48824, USA}

\author{Ya Zhang}
\email{zhangya1221@pku.edu.cn}
\affiliation{Department of Physics and State Key Laboratory of Nuclear Physics and Technology, Peking University, Beijing 100871, China}

\begin{abstract}
We propose to utilize the polarization information of the $Z$ bosons in $ZZ$ production, via the gluon-gluon fusion process $gg\to ZZ$, to probe the $Zt\bar{t}$ gauge coupling. The contribution of longitudinally polarized $Z$ bosons is sensitive to the axial-vector component ($a_t$) of the $Zt\bar{t}$ coupling. We demonstrate that the angular distribution of the charged lepton from $Z$ boson decays serves well for measuring the polarization of $Z$ bosons and the determination of $a_t$. We show that $ZZ$ production via the $gg\to ZZ$ process complement to $Zt\bar{t}$ and $tZj$ productions in measuring the $Zt\bar{t}$ coupling at hadron colliders. 
\end{abstract}

\maketitle

\noindent{\bf 1. Introduction.}

Top quark, the heaviest fermion in the Standard Model (SM), is commonly believed to be sensitive to new physics (NP) beyond the SM. The top quark often plays a key role in triggering electroweak symmetry breaking (EWSB) in many NP models, 
and as a result, the gauge couplings of top quarks, e.g. $Wtb$ and $Zt\bar{t}$, may largely deviate from the SM predictions~\cite{Martin:1997ns,Contino:2010rs,Bellazzini:2014yua,Panico:2015jxa,Csaki:2016kln}. The $Wtb$ couplings have been well measured in both the single top quark production and the top-quark decay~\cite{Chen:2005vr,Prasath:2014mfa,Cao:2015doa,Aguilar:2015vsa,Hioki:2015env,Buckley:2015lku,Zhang:2016omx,Birman:2016jhg,Jueid:2018wnj,Cao:2018ntd,Sun:2018byn,Cao:2019uor};
the $Zt\bar{t}$ coupling can be measured in $t\bar{t}Z$ and $tjZ$ productions~\cite{Baur:2004uw,Campbell:2013yla,Rontsch:2014cca,Cao:2015qta,Bylund:2016phk,Berger:2009hi,Degrande:2018fog,Martini:2019lsi} which are,   unfortunately, difficult to separately determine the vector and axial-vector components of the $Zt\bar{t}$ coupling.  
The chiral structure of the $Zt\bar{t}$ coupling would reveal the gauge structure of NP models~\cite{Richard:2014upa,Cao:2015qta}, therefore, measuring and distinguishing the vector and axial vector components of the $Zt\bar{t}$ coupling is in order.

In this work we explore the potential of measuring the $Zt\bar{t}$ coupling using the polarization information of the $Z$ bosons in $ZZ$ production, at the CERN Large Hadron Collider (LHC). The $Zt\bar{t}$ coupling contributes to $ZZ$ production through top-quark loop effects in the gluon fusion channel. 
The process, $gg\to ZZ$, has been used to constrain the Higgs boson width through the interference of box and triangle diagrams and  it has been shown to be sensitive to many NP effects~\cite{Caola:2013yja,Chen:2013waa,Campbell:2013una,Coleppa:2014qja,Gainer:2014hha,Azatov:2014jga,Englert:2014ffa,Li:2015jva,Englert:2015zra,Azatov:2016xik,Goncalves:2017iub,Lee:2018fxj,Goncalves:2018pkt,He:2019kgh}.
In particular, the polarization of the $Z$ boson pair highly depends on the $Zt\bar{t}$ coupling. 
The polarizations of $Z$ bosons in $ZZ$ pair production can be categorized as: TT (transverse-transverse), 
TL (transverse-longitudinal), and LL (longitudinal-longitudinal). 
\begin{figure}
	\includegraphics[scale=0.7]{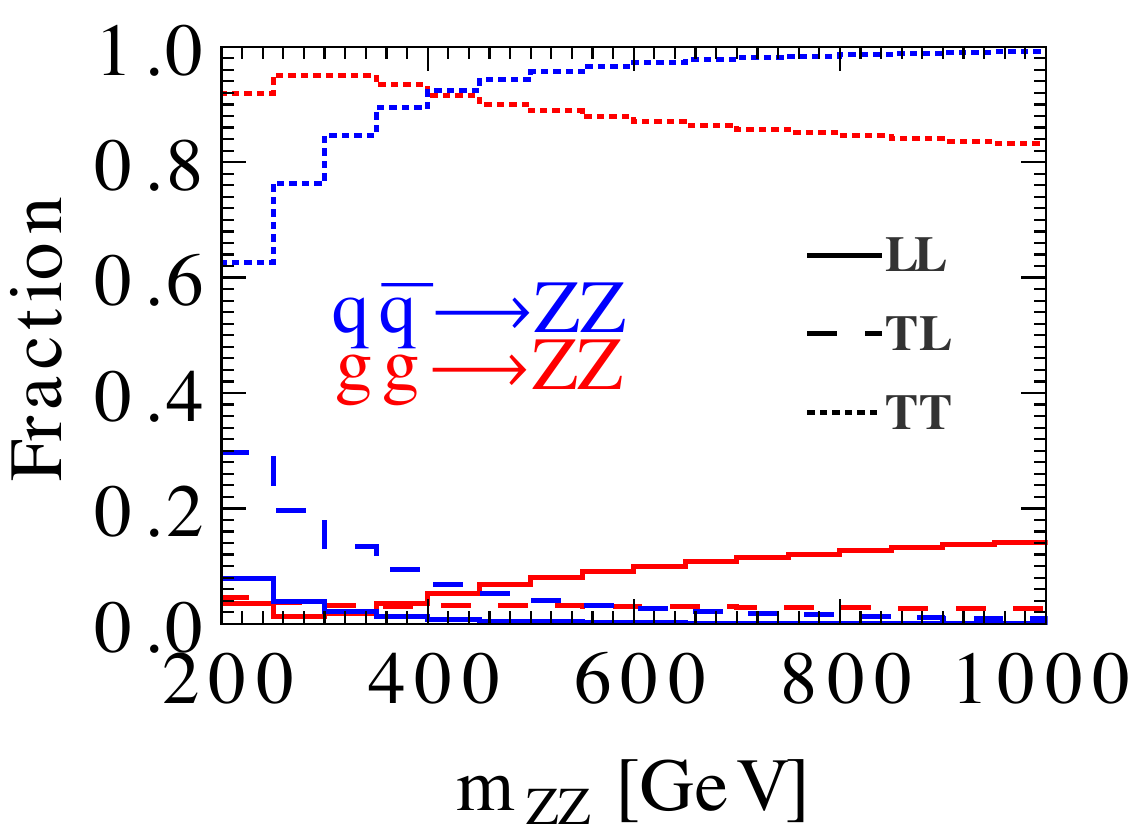}
	\caption{The fractions of three polarization modes in the processes of $gg\to ZZ$ and $q\bar{q}\to ZZ$ at the 13 TeV LHC.}
	\label{Fig:mzz}
\end{figure}
Figure~\ref{Fig:mzz} shows the fraction of the three polarization modes of $ZZ$ pairs in the processes of $gg\to ZZ$  (red) and $q\bar{q}\to ZZ$ (blue) at the 13 TeV LHC. The TT mode dominates in both production processes as a result of that, owing to the Goldstone boson equivalence theorem, the interaction of the longitudinal mode to light quarks is highly suppressed by the small mass of the light quarks.  
The suppression of the LL mode in the $gg\to ZZ$ channel arises from the cancellation between the box and triangle diagrams due to unitarity, and the cancellation is sensitive to the axial-vector coupling of $Zt\bar{t}$~\cite{Glover:1988rg}. In the  high energy limit, the contribution of top-quark loops to the LL mode is given by
\beq
M_{\pm,\pm,0,0}\sim \frac{m_t^2}{m_Z^2}\left(a_t^2-\frac{1}{4}\right)\left[\log^2\left(\frac{\hat{s}}{m_t^2}\right)-2i\pi\log\left(\frac{\hat{s}}{m_t^2}\right)\right],
\eeq 
where $a_t$ is the axial-vector component of the $Zt\bar{t}$ coupling, $m_{t}$ and $m_Z$ denotes the mass of the top quark and $Z$ boson, respectively. The subscript $+$, $-$ and $0$ denote the right-handed, left-handed and longitudinal polarization of the gluons or $Z$ bosons, respectively. In the SM,  $a_t=1/2$, and it yields a strong cancellation in the LL mode scattering. However, in the NP model the value of $a_t$ can deviate from its SM value, so that the above-mentioned cancellation is spoiled and 
the fraction of the LL mode contribution would be enhanced. 
Therefore, the polarization information of the $Z$ boson pairs in $ZZ$ production, via $gg\to ZZ$, can be used to probe the axial-vector coupling of $Zt\bar{t}$ interaction at hadron colliders.

\vspace{3mm}
\noindent{\bf 2. $ZZ$ production via Gluon fusion.}

Here, we consider the case that the NP effects modify only the four-dimensional $Zt\bar{t}$ coupling. 
The effective Lagrangian of the $Zt\bar{t}$ interaction is
\beq
\mathcal{L}=\frac{g_W}{2c_W}\bar{t}(v_t-a_t\gamma_5)\gamma_\mu t,
\eeq
where $g_W$ is the electroweak gauge coupling and $c_W$ is the cosine of the weak mixing angle $\theta_W$. In the SM,
\beq
v_t^{\rm SM}=\frac{1}{2}-\frac{2}{3} s_W^2=0.3526,\quad a_t^{\rm SM}=\frac{1}{2},
\eeq
where $s_W\equiv \sin\theta_W$. 
We calculate the helicity amplitudes of the channel $g(\ld_1)g(\ld_2)\to Z(\ld_3)Z(\ld_4)$ using FeynArts and  FeynCalc~\cite{Hahn:2000kx,Shtabovenko:2016sxi} where $\lambda_i$ labels the helicity of particle $i$.  
The contribution of the box diagram ($\square$) to each helicity amplitude can be parametrized as~\cite{Glover:1988rg},
\begin{align}
M^{\square}_{\ld_1,\ld_2,\ld_3,\ld_4}&=\left(v_t^2+a_t^2\right) A_{\ld_1,\ld_2,\ld_3,\ld_4} \nn\\
&+ \left(v_t^2-a_t^2\right)B_{\ld_1,\ld_2,\ld_3,\ld_4}\nn\\
&+ a_t^2 C_{\ld_1,\ld_2,\ld_3,\ld_4},
\label{eq:mbox}
\end{align}
where $\ld_i=\pm$ and $0$. Both $B_{\ld_1,\ld_2,\ld_3,\ld_4}$ and
$C_{\ld_1,\ld_2,\ld_3,\ld_4}$ vanish for (massless) light quark loops.  
In the limit of $\hat{s}=-\hat{t}/2=-\hat{u}/2 \gg m_t$, where $\hat{s},\hat{t}$ and $\hat{u}$ are the usual Mandelstam variables, the coefficients $A,B$ and $C$ are 
\begin{align}
&A\sim \text{Constant}, \nn\\
&B\sim0, \nn\\
&C_{\pm,\pm,0,0}\sim -\frac{m_{t}^{2}}{m_{Z}^{2}}\left[\log ^{2}\left(\frac{\hat{s}}{m_{t}^{2}} \right)-2i\pi\log\left(\frac{\hat{s}}{m_{t}^{2}} \right)\right].
\end{align}
Here, the constant in the coefficient $A$ is a combination of gauge couplings and loop factor.
Furthermore, The contribution of the triangle diagram ($\bigtriangleup$) to each helicity amplitude is 
\beq
M_{\pm,\pm,0,0}^{\bigtriangleup}\sim \frac{m_{t}^{2}}{4 m_{Z}^{2}}\left[\log ^{2}\left(\frac{\hat{s}}{m_{t}^{2}}\right)-2i\pi \log\left(\frac{\hat{s}}{m_{t}^{2}}\right)\right],
\eeq 
which cancels with the coefficient $C$ in the contribution of the box diagram $M^{\square}$ for each helicity amplitude.  
The sensitivity of the cancellation on $a_t$ can be understood from the fact  that the axial current is not conserved for the top quark, whose mass is at the weak scale. 

\begin{figure}
	\centering
	\includegraphics[scale=0.6]{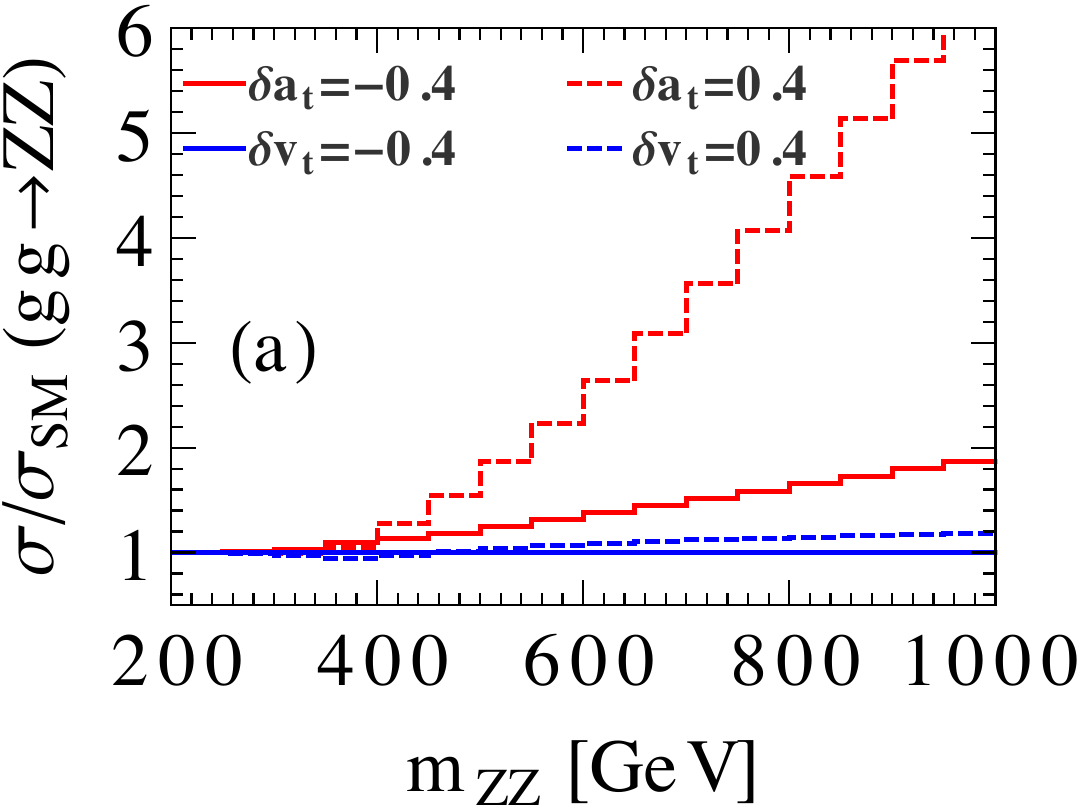}
	\includegraphics[scale=0.6]{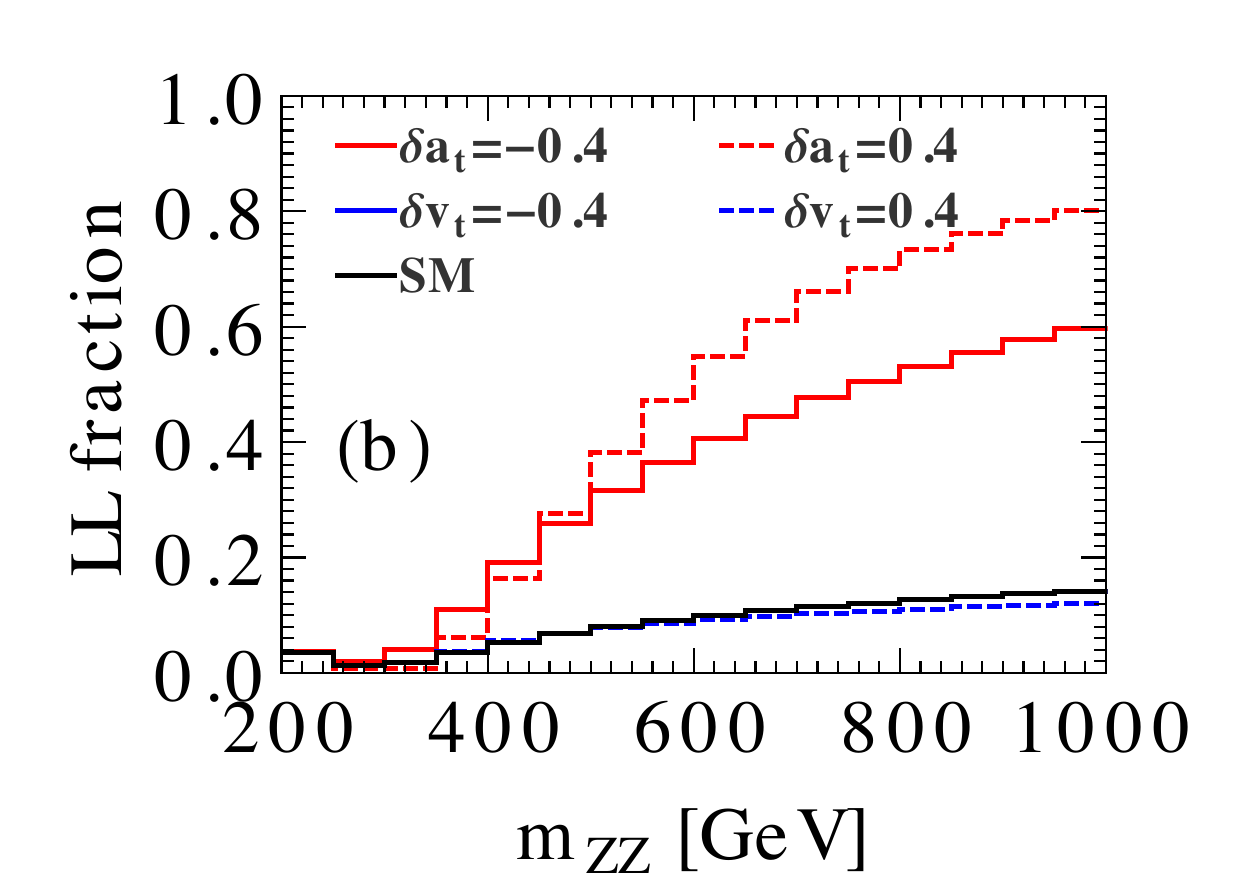}
         \includegraphics[scale=0.6]{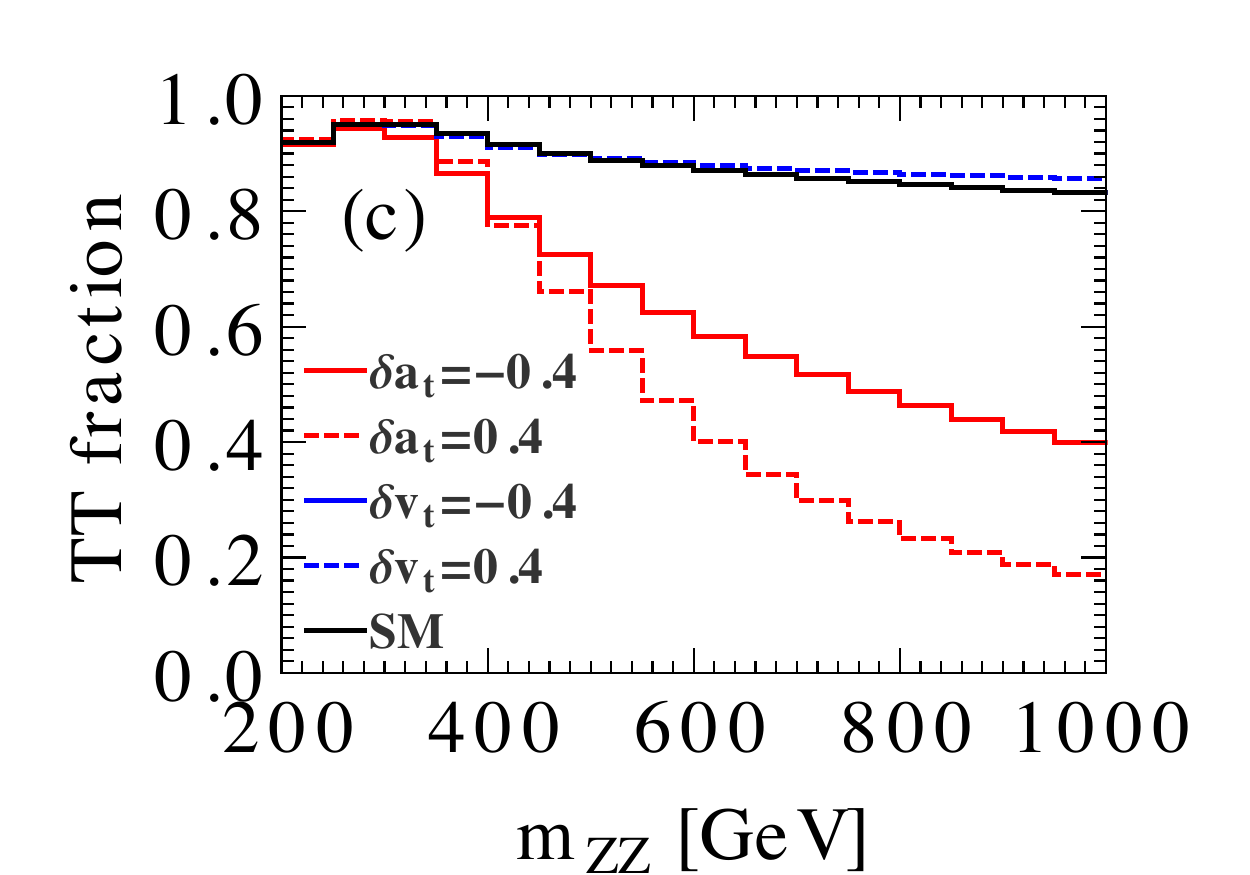}
	\caption{(a) The differential cross section of $gg\to ZZ$ for various $\delta v_t$'s and $\delta a_t$'s, normalized to the SM prediction, as a function of $m_{ZZ}$ at the 13 TeV LHC; the polarization fraction of the LL mode (b) and the TT mode (c). }
	\label{Fig:Rcs}
\end{figure}

Below, we consider the impact of the non-standard $Zt\bar{t}$ coupling to 
the differential cross sections of $gg\to ZZ$ by changing only one parameter at a time.
Both of the light and top quark loop contributions have been included in our numerical calculation. The light quark loop contribution gives the dominant contribution to the inclusive cross section, while it is only sensitive to the TT mode of the $ZZ$ pairs. Any deviation in the LL mode of the inclusive cross section, as studied in this work, can only come from the non-standard $Zt\bar{t}$ coupling. Furthermore, we have compared the result of our numerical calculations with that using the MadGraph5 code~\cite{Alwall:2014hca} and found excellent agreement.

Define $\delta v_t$ and $\delta a_t$  as the amount of deviation of the vector and axial-vector couplings from the SM values, i.e.,
\beq
\delta v_t=v_t-v_t^{\rm SM},\quad \delta a_t=a_t-a_t^{\rm SM}.
\eeq
Figure~\ref{Fig:Rcs}(a) shows the differential cross sections of $gg\to ZZ$, normalized to the SM prediction, as a function of  the invariant mass of the $Z$ boson pair ($m_{ZZ}$) for various $\delta v_t$'s and $\delta a_t$'s at the 13 TeV LHC.  Figure~\ref{Fig:Rcs}(b) and (c) show  the fraction of the LL and TT modes as a function of $m_{ZZ}$, respectively. The TL mode is not plotted as it is quite small in comparison with the LL and TT modes. The LL model is very sensitive to the anomalous $a_t$ coupling; for example, the contribution of the LL mode increases dramatically in the large $m_{ZZ}$ region for $\delta a_t=\pm 0.4$, cf.  the red solid and red dashed curves. On the other hand, the LL mode is not sensitive to $\delta v_t$. The fractions of the LL and TT modes are slightly altered for the choice of $\delta v_t=\pm 0.4$ and are very close to their fractions in the SM; cf. the blue and black curves. Therefore, the polarization information of the $Z$ bosons in $ZZ$ production can be utilized to provide a good probe of the anomalous $a_t$ coupling.

\begin{figure}
	\centering
	\includegraphics[scale=0.6]{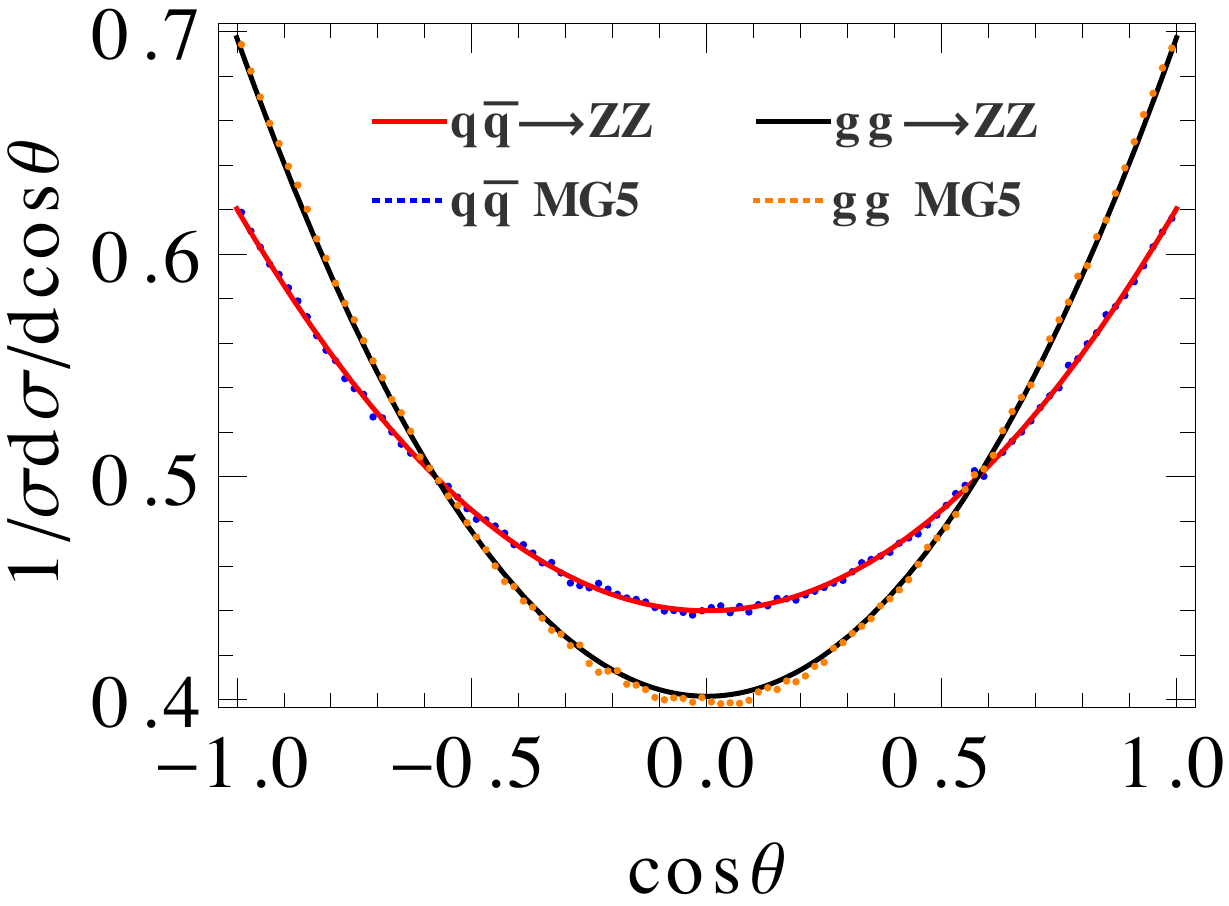}
	\caption{The $\cos\theta$ distribution, of the processes $gg, {q \bar q}\to ZZ\to 4\ell$ in the SM, where the solid and dotted curves denote the theory template and the MC simulation, respectively, without imposing any kinematic cut.}
	\label{Fig:theta}
\end{figure}

The polarization information of the final state $Z$ boson can be inferred from the angular distribution ($\cos\theta$) of the charged lepton in the rest frame of the $Z$ boson from which the charged lepton is emitted. The angular distributions for various polarization states of the $Z$ boson are given as 
\beq
\phi_L(\cos\theta)=\frac{3}{4}(1-\cos^2\theta),~ \phi_T(\cos\theta)=\frac{3}{8}(1+\cos^2\theta),
\eeq
where $\phi_{L}$ denotes a longitudinally polarized $Z$ boson,  and $\phi_{T}$ a transversely polarized $Z$. The angle $\theta$ is defined as the opening angle between the charged lepton three-momentum in the rest frame of the $Z$-boson and the $Z$-boson three-momentum in the center of mass frame of the $ZZ$ pair.

To determine the value of $a_t$, we compare 
the angular distributions ($\cos\theta$) predicted by the Monte Carlo (MC) simulation to the theory template obtained by the analytical calculation. 
The theory template of $\cos\theta$ distribution, of the processes 
$gg, {q \bar q}\to ZZ\to 4\ell$ with $\ell=e^\pm,\mu^\pm$, 
is approximated by multiplying the fraction of each polarization mode of the $ZZ$ boson pair with the corresponding $\phi_{L,T}$ distributions.  Though the spin correlation between the two final-state $Z$ bosons is not strictly maintained in this approximation, the prediction of the theory template (solid curves) on the $\cos\theta$ distribution, via either the $q \bar q$ or $gg$ scattering process, in the SM agrees very well with that obtained by the MC simulation (dashed curves), as clearly shown in Figure~\ref{Fig:theta} without imposing any kinematic cut.

\vspace{3mm}
\noindent{\bf 3. Collider simulation.}
 
 \begin{figure}
	\centering
	\includegraphics[scale=0.6]{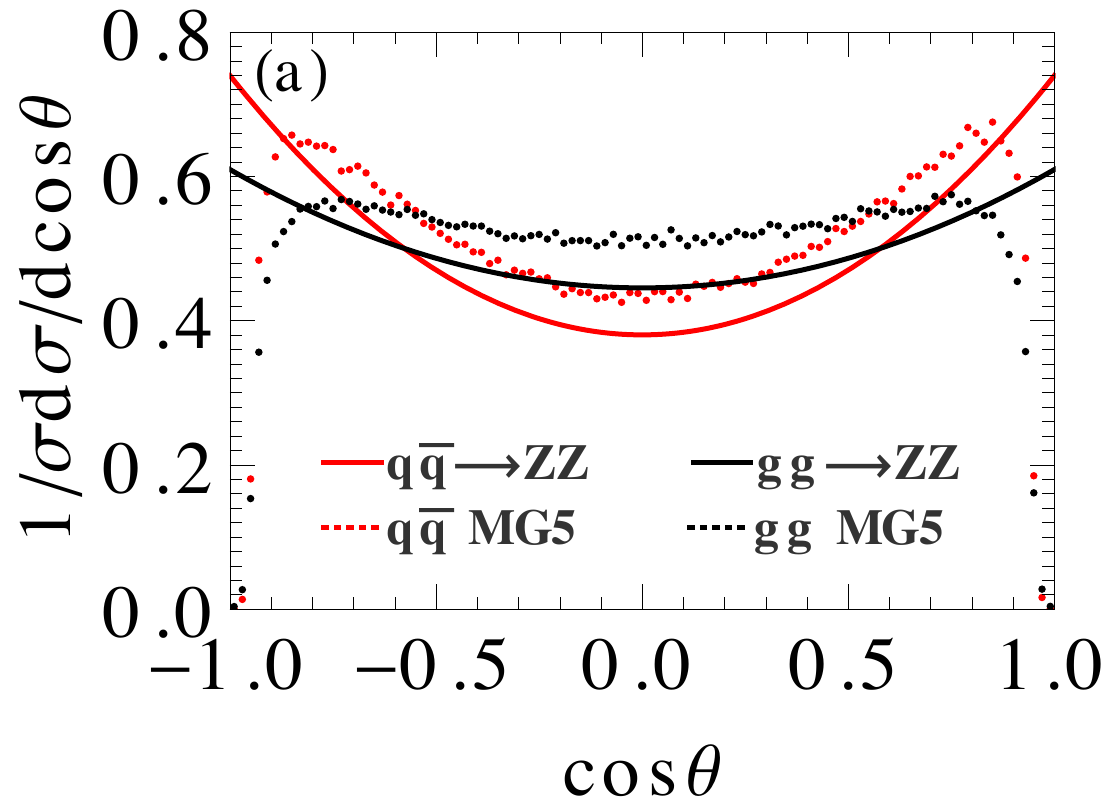}
	\includegraphics[scale=0.6]{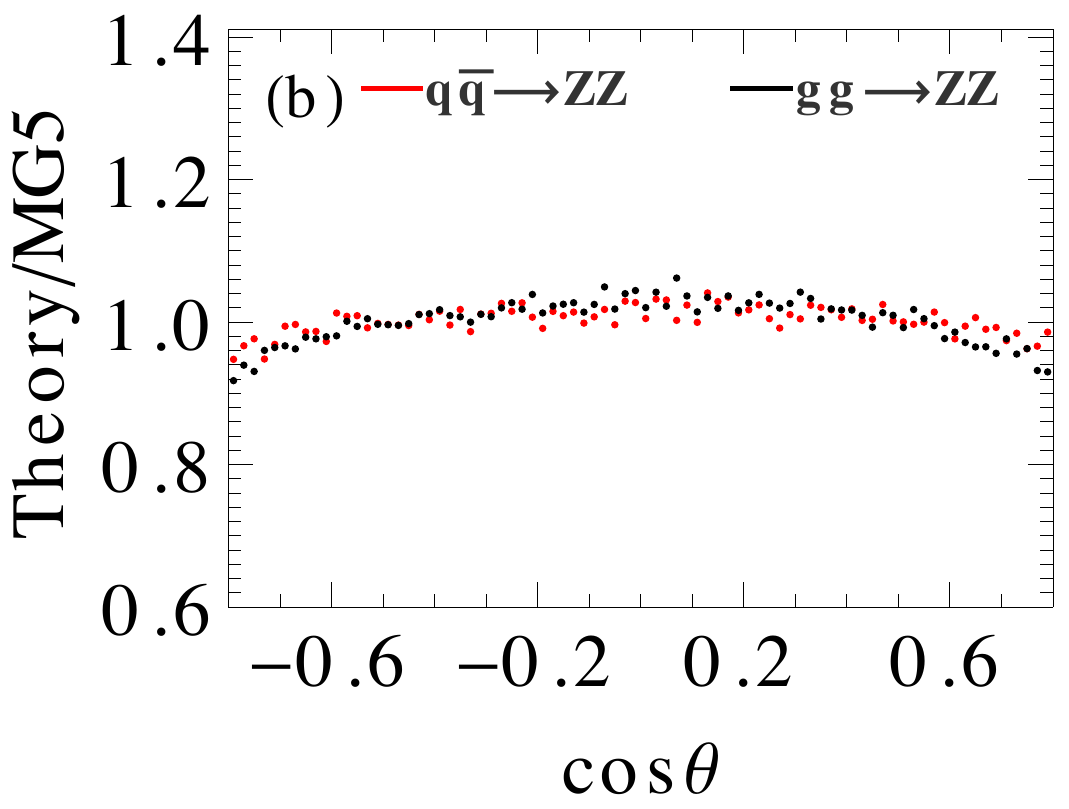}
	\caption{(a) The $\cos\theta$ distribution, of the processes $gg, {q \bar q}\to ZZ\to 4\ell$ in the SM, after imposing the kinematic cuts as described in the main text. The solid curves represents the theory template and the dotted curves denotes the MC simulation; (b) The ratio between the prediction of theory template and the MC data for $|\cos\theta| < 0.8$. }
	\label{Fig:theta2}
\end{figure}

 Next we perform a detailed Monte Carlo simulation to explore the potential of probing $a_t$ via the signal process $gg\to ZZ\to 4\ell$ at the 13 TeV LHC and a 100 TeV proton-proton (pp) hadron collider. 
 Its major background comes from the process $q\bar{q}\to ZZ$, while the other backgrounds are negligible~\cite{Aaboud:2018puo}. We generate both the signal and background events by MadGraph5~\cite{Alwall:2014hca} at the parton-level and pass events to PYTHIA~\cite{Sjostrand:2007gs} for showering and hadronization.
 The Delphes package is used to simulate the detector smearing effects~\cite{deFavereau:2013fsa}.  The QCD corrections are taken into account by introducing a constant $\kappa$ factor, i.e. $\kappa_{gg}=1.8$ and $\kappa_{q\bar{q}}=1.7$~\cite{Caola:2015psa,Cascioli:2014yka,Heinrich:2017bvg,Grazzini:2018owa,Kallweit:2018nyv,Agarwal:2019rag}.  At the analysis level, both  the signal and background events are required to pass the kinematic cuts: $|\eta_{\ell}|<2.5$ and $P_{T\ell} >15~{\rm GeV}$. We further require the invariant mass window cut for same flavor leptons as $80<m_{\ell\ell}<100~{\rm GeV}$ and demand $m_{4\ell}> 600~{\rm GeV}$ to enhance the LL mode. 
 
The kinematic cuts inevitably modify the lepton kinematics and the polarization fractions of the $ZZ$ bosons. 
In this study, we require $m_{ZZ}>600~{\rm GeV}$ and $|\eta_Z|<2$.
Figure~\ref{Fig:theta2}(a) displays the $\cos\theta$ distribution after imposing the kinematic cuts for the processes $gg\to ZZ$ (black) and $q\bar{q}\to ZZ$ (red). 
The shapes of the $\cos\theta$ distributions of the theory template agree with those of the MC simulation (labeled as MG5 in Fig.~\ref{Fig:theta2}) except near the edge region.  Note that the predictions of MG5 have included the effects from parton shower and detector level simulation. Focusing on the central region with  $|\cos\theta|<0.8$, we plot the ratio between the normalized theory template and the MC simulation in Fig~\ref{Fig:theta2}(b), which shows  good agreements between the two calculations. 
Hence, we applied the cut of $|\cos\theta|<0.8$ in the following  analysis, when using only the theory template predictions.

The total event number of the signal ($N_s$) and background ($N_b$) processes are
\begin{align}
N_{s}=&\sigma(gg\to ZZ)\times 4{\rm Br}^2\times \epsilon_{\rm cut}^{g}\times\mathcal{L}_{\rm int},\nn\\
N_{b}=&\sigma(q\bar{q}\to ZZ)\times 4{\rm Br}^2\times \epsilon_{\rm cut}^{q}\times\mathcal{L}_{\rm int},
\end{align}
where $\epsilon_{\rm cut}^{g,q}$ is the cut efficiency for the signal and background process, respectively. $\mathcal{L}_{\rm int}$ is the integrated luminosity, and 
\begin{align}
{\rm Br}\equiv {\rm Br}(Z\to e^+e^-)={\rm Br}(Z\to \mu^+\mu^-). 
\end{align}
In the SM (with $\delta v_t=\delta a_v=0$),  the total cross section of the signal ($\sigma_s$) and background ($\sigma_b$) processes are,
\begin{align}
&\sigma_s=\sigma(gg\to ZZ)\times 4{\rm Br}^2\times \epsilon_{\rm cut}^{g}\simeq0.032~{\rm fb},\nn\\
&\sigma_b=\sigma(q\bar{q}\to ZZ)\times 4{\rm Br}^2\times \epsilon_{\rm cut}^{q}\simeq0.2~{\rm fb}
\end{align}
at the 13 TeV LHC, while at a 100 TeV pp collider
\begin{align}
&\sigma_s=\sigma(gg\to ZZ)\times 4{\rm Br}^2\times \epsilon_{\rm cut}^{g}\simeq 1.26~{\rm fb},\nn\\
&\sigma_b=\sigma(q\bar{q}\to ZZ)\times 4{\rm Br}^2\times \epsilon_{\rm cut}^{q}\simeq 1.72~{\rm fb}.
\end{align}
There are roughly about 700 and 9000 events of $ZZ$ pairs produced at the 13~TeV LHC and a 100~TeV pp collider with an integrated luminosity of $3000~{\rm fb}^{-1}$.

For probing the $Zt\bar{t}$ coupling, we divide the $|\cos\theta|$ distribution into 8 bins and use the binned likelihood function to estimate the sensitivity for the hypothesis of NP with a non-vanishing $(\delta v_t,\delta a_t)$ against the hypothesis of the SM coupling~\cite{Cowan:2010js},
\beq
L(\delta v_t,\delta a_t)=\prod_{i=1}^{N_{\rm bin}}\frac{(s_i(\delta v_t,\delta a_t)+b_i)^{n_i}}{n_i!}e^{-s_i(\delta v_t,\delta a_t)-b_i},
\eeq
where $n_i$ denotes the number of observed events in the $i$th bin,  $b_i$ the number of background events, and $s_i(\delta v_t,\delta a_t)$ the number of signal events with the anomalous coupling $(\delta v_t,\delta a_t)$. The observed event is assumed to be $n_i=b_i+s_i(0,0)$.
The numbers of the signal events ($s_i$) and the background events ($b_i$) in each bin are determined by the total cross section, the fraction of polarization modes of the $Z$ boson pair and the $\phi_{L,T}$ functions, i.e.,

\beq
s_i(\delta v_t,\delta a_t),b_i=F_N^{g,q}N_{s,b}\int_i d \cos\theta \left[R_L^{g,q} \phi_L+(1-R_L^{g,q})\phi_T\right],
\eeq
where $R_L^{g,q}$ is the fraction of a longitudinal polarized $Z$ boson, which decays into a pair of electron or muon leptons, in the scattering processes $gg\to ZZ$ and $q\bar{q}\to ZZ$, respectively.  Here, $F_N^{g,q}$ is the normalization factor to ensure that 
$F_N^{g,q} \int_{-0.8}^{0.8} d \cos\theta \left[R_L^{g,q} \phi_L+(1-R_L^{g,q})\phi_T\right] =1.$
Explicitly, $F_N^{g,q}=1/(0.728+0.216 R_L^{g,q})$. 
Define the likelihood function ratio as following,
\beq
q^2=-2\frac{L(\delta v_t,\delta a_t)}{L(0,0)},
\eeq
which describes the exclusion of the hypothesis of NP with non-zero $(\delta v_t,\delta a_t)$ versus the hypothesis of SM at the $q$-sigma ($q\sigma$) level.

\begin{figure}
	\centering
	\includegraphics[scale=0.6]{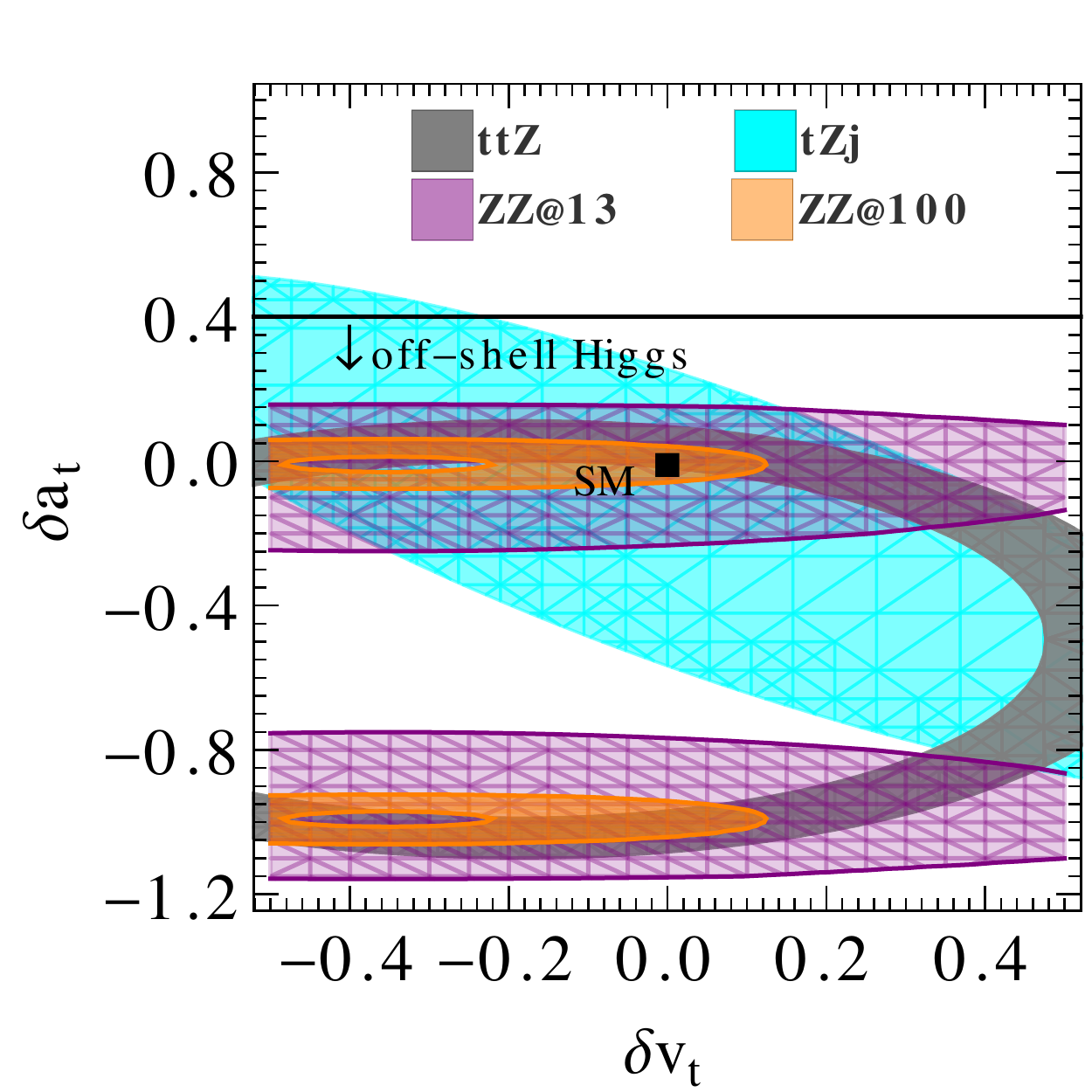}
	\caption{The parameter space of $(\delta v_t,\delta a_t)$, to be constrained by the measurement of $ZZ$ production, at the $2\sigma$ level at the 13 TeV LHC (purple region) and a 100 TeV pp hadron collider (orange region), respectively,  with an integrated luminosity of $3000~\rm{fb}^{-1}$. The gray region represents the present constraint from the $Zt\bar{t}$ production~\cite{Aaboud:2019njj,CMS:2019too}, and the cyan region from the $tZj$ production~\cite{Sirunyan:2018zgs,Aad:2020wog} at the 13~TeV LHC. 
}
	\label{Fig:dis}
\end{figure}

Figure~\ref{Fig:dis} displays the projected regions of the parameter space in which $(\delta v_t,\delta a_t)$ can be measured at the $2\sigma$ level, at the 13 TeV LHC and a 100 TeV pp hadron collider with an integrated luminosity of $3000~\rm{fb}^{-1}$. The cyan and gray regions denote the constraints provided by the measurement of $tZj$ \cite{Aad:2020wog,Sirunyan:2018zgs} and $Zt\bar{t}$ \cite{CMS:2019too,Aaboud:2019njj} productions at the 13~TeV LHC, respectively. The horizontal black line represents the upper limit of $\delta a_t$ derived from the strength of the off-shell Higgs-boson signal in $ZZ$ production~\cite{Aaboud:2018puo}.
The purple region denotes the projected parameter space obtained from measuring the degree of polarization of the $Z$ bosons in $gg \to ZZ$ production, at the $2\sigma$ level, at the 13 TeV LHC, while the orange region is the projected parameter space for a 100 TeV pp collider.  

It is evident that 
the measurement of $Zt\bar{t}$ production, as compared to $ZZ$ and $tZj$ productions, yields the strongest constraint on values of $\delta v_t$ and $\delta a_t$ at the 13 TeV LHC.
However, the drawback of this measurement is that the bounded region contains a degeneracy of $\delta a_t$ and $\delta v_t$, i.e. 
\begin{equation}
0.77\leq 3.05\left(\delta a_t+0.5\right)^2+1.71\left(\delta v_t+0.19\right)^2\leq 1.14.
\end{equation}
Taking into account the $tZj$ production can partially resolve the degeneracy, found in analyzing the $Zt\bar{t}$ events. The $ZZ$ production is sensitive only to $a_t$, and it alone yields a twofold constraint 
$\delta a_t \in [-0.25,0.15] \cup [-1.16,-0.75]$ at the 13 TeV LHC, and $\delta a_t\in [-0.08,0.06]\cup [-1.00,-0.92]$ at a 100 TeV pp collider, cf. the two purple and orange regions. 
With a larger data sample in the future runs of the LHC and a 100 TeV pp collider, it is possible to precisely determine first the axial-vector component $a_t$, and then the vector component $v_t$ of the  $Zt\bar{t}$ coupling. 
The measurement of $tZj$ production is particularly important for the determination of its vector component from the combined analysis.  It was shown in Ref.~\cite{Mangano:2016jyj,Vos:2017jxz} that at a 100 TeV pp collider, the measurement of the $Zt\bar{t}$ coupling could be further improved by studying the $t\bar{t}Z$ and $tjZ$ production cross sections and its uncertainty can be controlled to within a few percent level.

Before closing this section, we would like to compare our findings, derived from studying the polarization state of the produced $ZZ$ pairs from $gg$ fusion, with that in the literature, obtained from studying the inclusive production rates alone. 
In Ref.~\cite{Azatov:2016xik}, it was concluded that $a_t$ can be constrained as $\delta a_t/a_t\in [-0.42,0.35]$ by measuring the $gg\to ZZ$ inclusive cross section at the 14 TeV LHC with an integrated luminosity of $3~\rm{ab}^{-1}$. 
With $a_t^{\rm SM}=0.5$, the result of our analysis, invoking the polarization information of the final state $ZZ$ pairs,  yields $\delta a_t/a_t\in [-0.44,0.3]$, though it is for a 13 TeV LHC. It appears that our result only slightly improve the sensitivity of this production channel to the measurement of the $Zt\bar{t}$ coupling. However, the main point made and demonstrated in this work is that the LL mode of the $ZZ$ pair production is sensitive to the anomalous $a_t$ (but not $v_t$) coupling of top quark to $Z$ boson. Hence, it can be used to help disentangle the contributions of both $a_t$ and $v_t$ couplings in the total inclusive cross section measurement. Moreover, the result presented in this work could potentially be improved if one utilizes advanced technologies such as Boosted Decision Tree or Multi-Variable Analysis~\cite{Lee:2018fxj,Lee:2019nhm}, which is however beyond the scope of this work.

\vspace{3mm}
\noindent{\bf 4. Summary.}

We propose to measure the axial-vector component of the $Zt\bar{t}$ coupling by utilizing the polarization information of the $Z$ bosons in the process $gg\to ZZ$, at the 13~TeV LHC and a 100~TeV proton-proton collider. When the final-state $Z$-bosons are both longitudinally polarized, the cross section for $gg\to ZZ$ is sensitive to the axial-vector coupling $a_t$, because the axial current is not conserved for massive top quarks. 
We demonstrate that the fraction of longitudinal-longitudinal (LL) mode increases with a non-vanishing anomalous  coupling $a_t$, when the invariant mass of the $Z$-boson pair becomes larger.  From the angular distribution of the charged leptons from the $Z$-boson decay, one can determine the polarization of the $Z$ bosons and in turn to probe the anomalous $a_t$ coupling, regardless of the value of the vector component ($v_t$) of the $Zt\bar{t}$ coupling. By comparing the theory template and Monte Carlo simulation, we find the parameter space of $\delta a_t\equiv a_t-a_t^{\rm SM}$ which can be probed 
at the $2\sigma$ level, from 
the measurement of $ZZ$ production at hadron colliders. 
It is  $\delta a_t \in [-0.25,0.15] \cup [-1.16,-0.75]$ at the 13 TeV LHC, and  $\delta a_t\in [-0.08,0.06]\cup [-1.00,-0.92]$ at a 100~TeV pp collider. We emphasize that the $ZZ$ production is complementary to the $Zt\bar{t}$ and $tZj$ productions in the measurement of the $Zt\bar{t}$ coupling. 
 
\noindent{\bf Acknowledgments.}
BY thank  Yandong Liu, Zhuoni Qian and  Ling-Xiao Xu for helpful discussion. 
QHC and YZ are supported in part by the National Science Foundation of China 
under Grant Nos. 11725520, 11675002 and 11635001. 
BY was supported by the U.S. Department of Energy through the Office of Science, Office of Nuclear Physics under Contract DE-AC52-06NA25396 and by an Early Career Research Award (C. Lee).
C.-P. Yuan was supported  by the U.S. National
Science Foundation under Grant No. PHY-1719914.
C.-P. Yuan is also grateful for the support from the Wu-Ki Tung endowed chair
in particle physics.

\bibliographystyle{apsrev}
\bibliography{reference}

\begin{thebibliography}{61}
\expandafter\ifx\csname natexlab\endcsname\relax\def\natexlab#1{#1}\fi
\expandafter\ifx\csname bibnamefont\endcsname\relax
  \def\bibnamefont#1{#1}\fi
\expandafter\ifx\csname bibfnamefont\endcsname\relax
  \def\bibfnamefont#1{#1}\fi
\expandafter\ifx\csname citenamefont\endcsname\relax
  \def\citenamefont#1{#1}\fi
\expandafter\ifx\csname url\endcsname\relax
  \def\url#1{\texttt{#1}}\fi
\expandafter\ifx\csname urlprefix\endcsname\relax\def\urlprefix{URL }\fi
\providecommand{\bibinfo}[2]{#2}
\providecommand{\eprint}[2][]{\url{#2}}

\bibitem[{\citenamefont{Martin}(1997)}]{Martin:1997ns}
\bibinfo{author}{\bibfnamefont{S.~P.} \bibnamefont{Martin}}, pp.
  \bibinfo{pages}{1--98} (\bibinfo{year}{1997}), \bibinfo{note}{[Adv. Ser.
  Direct. High Energy Phys.18,1(1998)]}, \eprint{hep-ph/9709356}.

\bibitem[{\citenamefont{Contino}(2011)}]{Contino:2010rs}
\bibinfo{author}{\bibfnamefont{R.}~\bibnamefont{Contino}}, in
  \emph{\bibinfo{booktitle}{{Physics of the large and the small, TASI 09,
  proceedings of the Theoretical Advanced Study Institute in Elementary
  Particle Physics, Boulder, Colorado, USA, 1-26 June 2009}}}
  (\bibinfo{year}{2011}), pp. \bibinfo{pages}{235--306}, \eprint{1005.4269}.

\bibitem[{\citenamefont{Bellazzini et~al.}(2014)\citenamefont{Bellazzini,
  Csaki, and Serra}}]{Bellazzini:2014yua}
\bibinfo{author}{\bibfnamefont{B.}~\bibnamefont{Bellazzini}},
  \bibinfo{author}{\bibfnamefont{C.}~\bibnamefont{Csaki}}, \bibnamefont{and}
  \bibinfo{author}{\bibfnamefont{J.}~\bibnamefont{Serra}},
  \bibinfo{journal}{Eur. Phys. J.} \textbf{\bibinfo{volume}{C74}},
  \bibinfo{pages}{2766} (\bibinfo{year}{2014}), \eprint{1401.2457}.

\bibitem[{\citenamefont{Panico and Wulzer}(2016)}]{Panico:2015jxa}
\bibinfo{author}{\bibfnamefont{G.}~\bibnamefont{Panico}} \bibnamefont{and}
  \bibinfo{author}{\bibfnamefont{A.}~\bibnamefont{Wulzer}},
  \bibinfo{journal}{Lect. Notes Phys.} \textbf{\bibinfo{volume}{913}},
  \bibinfo{pages}{pp.1} (\bibinfo{year}{2016}), \eprint{1506.01961}.

\bibitem[{\citenamefont{Csaki and Tanedo}(2015)}]{Csaki:2016kln}
\bibinfo{author}{\bibfnamefont{C.}~\bibnamefont{Csaki}} \bibnamefont{and}
  \bibinfo{author}{\bibfnamefont{P.}~\bibnamefont{Tanedo}}, in
  \emph{\bibinfo{booktitle}{{Proceedings, 2013 European School of High-Energy
  Physics (ESHEP 2013): Paradfurdo, Hungary, June 5-18, 2013}}}
  (\bibinfo{year}{2015}), pp. \bibinfo{pages}{169--268}, \eprint{1602.04228}.

\bibitem[{\citenamefont{Chen et~al.}(2005)\citenamefont{Chen, Larios, and
  Yuan}}]{Chen:2005vr}
\bibinfo{author}{\bibfnamefont{C.-R.} \bibnamefont{Chen}},
  \bibinfo{author}{\bibfnamefont{F.}~\bibnamefont{Larios}}, \bibnamefont{and}
  \bibinfo{author}{\bibfnamefont{C.~P.} \bibnamefont{Yuan}},
  \bibinfo{journal}{Phys. Lett.} \textbf{\bibinfo{volume}{B631}},
  \bibinfo{pages}{126} (\bibinfo{year}{2005}), \eprint{hep-ph/0503040}.

\bibitem[{\citenamefont{Prasath~V et~al.}(2015)\citenamefont{Prasath~V,
  Godbole, and Rindani}}]{Prasath:2014mfa}
\bibinfo{author}{\bibfnamefont{A.}~\bibnamefont{Prasath~V}},
  \bibinfo{author}{\bibfnamefont{R.~M.} \bibnamefont{Godbole}},
  \bibnamefont{and} \bibinfo{author}{\bibfnamefont{S.~D.}
  \bibnamefont{Rindani}}, \bibinfo{journal}{Eur. Phys. J.}
  \textbf{\bibinfo{volume}{C75}}, \bibinfo{pages}{402} (\bibinfo{year}{2015}),
  \eprint{1405.1264}.

\bibitem[{\citenamefont{Cao et~al.}(2017)\citenamefont{Cao, Yan, Yu, and
  Zhang}}]{Cao:2015doa}
\bibinfo{author}{\bibfnamefont{Q.-H.} \bibnamefont{Cao}},
  \bibinfo{author}{\bibfnamefont{B.}~\bibnamefont{Yan}},
  \bibinfo{author}{\bibfnamefont{J.-H.} \bibnamefont{Yu}}, \bibnamefont{and}
  \bibinfo{author}{\bibfnamefont{C.}~\bibnamefont{Zhang}},
  \bibinfo{journal}{Chin. Phys.} \textbf{\bibinfo{volume}{C41}},
  \bibinfo{pages}{063101} (\bibinfo{year}{2017}), \eprint{1504.03785}.

\bibitem[{\citenamefont{Romero~Aguilar
  et~al.}(2015)\citenamefont{Romero~Aguilar, Bouzas, and
  Larios}}]{Aguilar:2015vsa}
\bibinfo{author}{\bibfnamefont{R.}~\bibnamefont{Romero~Aguilar}},
  \bibinfo{author}{\bibfnamefont{A.~O.} \bibnamefont{Bouzas}},
  \bibnamefont{and} \bibinfo{author}{\bibfnamefont{F.}~\bibnamefont{Larios}},
  \bibinfo{journal}{Phys. Rev.} \textbf{\bibinfo{volume}{D92}},
  \bibinfo{pages}{114009} (\bibinfo{year}{2015}), \eprint{1509.06431}.

\bibitem[{\citenamefont{Hioki and Ohkuma}(2016)}]{Hioki:2015env}
\bibinfo{author}{\bibfnamefont{Z.}~\bibnamefont{Hioki}} \bibnamefont{and}
  \bibinfo{author}{\bibfnamefont{K.}~\bibnamefont{Ohkuma}},
  \bibinfo{journal}{Phys. Lett.} \textbf{\bibinfo{volume}{B752}},
  \bibinfo{pages}{128} (\bibinfo{year}{2016}), \eprint{1511.03437}.

\bibitem[{\citenamefont{Buckley et~al.}(2016)\citenamefont{Buckley, Englert,
  Ferrando, Miller, Moore, Russell, and White}}]{Buckley:2015lku}
\bibinfo{author}{\bibfnamefont{A.}~\bibnamefont{Buckley}},
  \bibinfo{author}{\bibfnamefont{C.}~\bibnamefont{Englert}},
  \bibinfo{author}{\bibfnamefont{J.}~\bibnamefont{Ferrando}},
  \bibinfo{author}{\bibfnamefont{D.~J.} \bibnamefont{Miller}},
  \bibinfo{author}{\bibfnamefont{L.}~\bibnamefont{Moore}},
  \bibinfo{author}{\bibfnamefont{M.}~\bibnamefont{Russell}}, \bibnamefont{and}
  \bibinfo{author}{\bibfnamefont{C.~D.} \bibnamefont{White}},
  \bibinfo{journal}{JHEP} \textbf{\bibinfo{volume}{04}}, \bibinfo{pages}{015}
  (\bibinfo{year}{2016}), \eprint{1512.03360}.

\bibitem[{\citenamefont{Zhang}(2016)}]{Zhang:2016omx}
\bibinfo{author}{\bibfnamefont{C.}~\bibnamefont{Zhang}},
  \bibinfo{journal}{Phys. Rev. Lett.} \textbf{\bibinfo{volume}{116}},
  \bibinfo{pages}{162002} (\bibinfo{year}{2016}), \eprint{1601.06163}.

\bibitem[{\citenamefont{Birman et~al.}(2016)\citenamefont{Birman, Deliot,
  Fiolhais, Onofre, and Pease}}]{Birman:2016jhg}
\bibinfo{author}{\bibfnamefont{J.~L.} \bibnamefont{Birman}},
  \bibinfo{author}{\bibfnamefont{F.}~\bibnamefont{Deliot}},
  \bibinfo{author}{\bibfnamefont{M.~C.~N.} \bibnamefont{Fiolhais}},
  \bibinfo{author}{\bibfnamefont{A.}~\bibnamefont{Onofre}}, \bibnamefont{and}
  \bibinfo{author}{\bibfnamefont{C.~M.} \bibnamefont{Pease}},
  \bibinfo{journal}{Phys. Rev.} \textbf{\bibinfo{volume}{D93}},
  \bibinfo{pages}{113021} (\bibinfo{year}{2016}), \eprint{1605.02679}.

\bibitem[{\citenamefont{Jueid}(2018)}]{Jueid:2018wnj}
\bibinfo{author}{\bibfnamefont{A.}~\bibnamefont{Jueid}},
  \bibinfo{journal}{Phys. Rev.} \textbf{\bibinfo{volume}{D98}},
  \bibinfo{pages}{053006} (\bibinfo{year}{2018}), \eprint{1805.07763}.

\bibitem[{\citenamefont{Cao et~al.}(2018)\citenamefont{Cao, Sun, Yan, Yuan, and
  Yuan}}]{Cao:2018ntd}
\bibinfo{author}{\bibfnamefont{Q.-H.} \bibnamefont{Cao}},
  \bibinfo{author}{\bibfnamefont{P.}~\bibnamefont{Sun}},
  \bibinfo{author}{\bibfnamefont{B.}~\bibnamefont{Yan}},
  \bibinfo{author}{\bibfnamefont{C.~P.} \bibnamefont{Yuan}}, \bibnamefont{and}
  \bibinfo{author}{\bibfnamefont{F.}~\bibnamefont{Yuan}},
  \bibinfo{journal}{Phys. Rev.} \textbf{\bibinfo{volume}{D98}},
  \bibinfo{pages}{054032} (\bibinfo{year}{2018}), \eprint{1801.09656}.

\bibitem[{\citenamefont{Sun et~al.}(2019)\citenamefont{Sun, Yan, and
  Yuan}}]{Sun:2018byn}
\bibinfo{author}{\bibfnamefont{P.}~\bibnamefont{Sun}},
  \bibinfo{author}{\bibfnamefont{B.}~\bibnamefont{Yan}}, \bibnamefont{and}
  \bibinfo{author}{\bibfnamefont{C.~P.} \bibnamefont{Yuan}},
  \bibinfo{journal}{Phys. Rev.} \textbf{\bibinfo{volume}{D99}},
  \bibinfo{pages}{034008} (\bibinfo{year}{2019}), \eprint{1811.01428}.

\bibitem[{\citenamefont{Cao et~al.}(2019)\citenamefont{Cao, Sun, Yan, Yuan, and
  Yuan}}]{Cao:2019uor}
\bibinfo{author}{\bibfnamefont{Q.-H.} \bibnamefont{Cao}},
  \bibinfo{author}{\bibfnamefont{P.}~\bibnamefont{Sun}},
  \bibinfo{author}{\bibfnamefont{B.}~\bibnamefont{Yan}},
  \bibinfo{author}{\bibfnamefont{C.~P.} \bibnamefont{Yuan}}, \bibnamefont{and}
  \bibinfo{author}{\bibfnamefont{F.}~\bibnamefont{Yuan}}
  (\bibinfo{year}{2019}), \eprint{1902.09336}.

\bibitem[{\citenamefont{Baur et~al.}(2005)\citenamefont{Baur, Juste, Orr, and
  Rainwater}}]{Baur:2004uw}
\bibinfo{author}{\bibfnamefont{U.}~\bibnamefont{Baur}},
  \bibinfo{author}{\bibfnamefont{A.}~\bibnamefont{Juste}},
  \bibinfo{author}{\bibfnamefont{L.~H.} \bibnamefont{Orr}}, \bibnamefont{and}
  \bibinfo{author}{\bibfnamefont{D.}~\bibnamefont{Rainwater}},
  \bibinfo{journal}{Phys. Rev.} \textbf{\bibinfo{volume}{D71}},
  \bibinfo{pages}{054013} (\bibinfo{year}{2005}), \eprint{hep-ph/0412021}.

\bibitem[{\citenamefont{Campbell et~al.}(2013)\citenamefont{Campbell, Ellis,
  and Rontsch}}]{Campbell:2013yla}
\bibinfo{author}{\bibfnamefont{J.}~\bibnamefont{Campbell}},
  \bibinfo{author}{\bibfnamefont{R.~K.} \bibnamefont{Ellis}}, \bibnamefont{and}
  \bibinfo{author}{\bibfnamefont{R.}~\bibnamefont{Rontsch}},
  \bibinfo{journal}{Phys. Rev. D} \textbf{\bibinfo{volume}{87}},
  \bibinfo{pages}{114006} (\bibinfo{year}{2013}), \eprint{1302.3856}.

\bibitem[{\citenamefont{Rontsch and Schulze}(2014)}]{Rontsch:2014cca}
\bibinfo{author}{\bibfnamefont{R.}~\bibnamefont{Rontsch}} \bibnamefont{and}
  \bibinfo{author}{\bibfnamefont{M.}~\bibnamefont{Schulze}},
  \bibinfo{journal}{JHEP} \textbf{\bibinfo{volume}{07}}, \bibinfo{pages}{091}
  (\bibinfo{year}{2014}), \bibinfo{note}{[Erratum: JHEP09,132(2015)]},
  \eprint{1404.1005}.

\bibitem[{\citenamefont{Cao and Yan}(2015)}]{Cao:2015qta}
\bibinfo{author}{\bibfnamefont{Q.-H.} \bibnamefont{Cao}} \bibnamefont{and}
  \bibinfo{author}{\bibfnamefont{B.}~\bibnamefont{Yan}},
  \bibinfo{journal}{Phys. Rev.} \textbf{\bibinfo{volume}{D92}},
  \bibinfo{pages}{094018} (\bibinfo{year}{2015}), \eprint{1507.06204}.

\bibitem[{\citenamefont{Bessidskaia~Bylund
  et~al.}(2016)\citenamefont{Bessidskaia~Bylund, Maltoni, Tsinikos, Vryonidou,
  and Zhang}}]{Bylund:2016phk}
\bibinfo{author}{\bibfnamefont{O.}~\bibnamefont{Bessidskaia~Bylund}},
  \bibinfo{author}{\bibfnamefont{F.}~\bibnamefont{Maltoni}},
  \bibinfo{author}{\bibfnamefont{I.}~\bibnamefont{Tsinikos}},
  \bibinfo{author}{\bibfnamefont{E.}~\bibnamefont{Vryonidou}},
  \bibnamefont{and} \bibinfo{author}{\bibfnamefont{C.}~\bibnamefont{Zhang}},
  \bibinfo{journal}{JHEP} \textbf{\bibinfo{volume}{05}}, \bibinfo{pages}{052}
  (\bibinfo{year}{2016}), \eprint{1601.08193}.

\bibitem[{\citenamefont{Berger et~al.}(2009)\citenamefont{Berger, Cao, and
  Low}}]{Berger:2009hi}
\bibinfo{author}{\bibfnamefont{E.~L.} \bibnamefont{Berger}},
  \bibinfo{author}{\bibfnamefont{Q.-H.} \bibnamefont{Cao}}, \bibnamefont{and}
  \bibinfo{author}{\bibfnamefont{I.}~\bibnamefont{Low}},
  \bibinfo{journal}{Phys. Rev.} \textbf{\bibinfo{volume}{D80}},
  \bibinfo{pages}{074020} (\bibinfo{year}{2009}), \eprint{0907.2191}.

\bibitem[{\citenamefont{Degrande et~al.}(2018)\citenamefont{Degrande, Maltoni,
  Mimasu, Vryonidou, and Zhang}}]{Degrande:2018fog}
\bibinfo{author}{\bibfnamefont{C.}~\bibnamefont{Degrande}},
  \bibinfo{author}{\bibfnamefont{F.}~\bibnamefont{Maltoni}},
  \bibinfo{author}{\bibfnamefont{K.}~\bibnamefont{Mimasu}},
  \bibinfo{author}{\bibfnamefont{E.}~\bibnamefont{Vryonidou}},
  \bibnamefont{and} \bibinfo{author}{\bibfnamefont{C.}~\bibnamefont{Zhang}},
  \bibinfo{journal}{JHEP} \textbf{\bibinfo{volume}{10}}, \bibinfo{pages}{005}
  (\bibinfo{year}{2018}), \eprint{1804.07773}.

\bibitem[{\citenamefont{Martini and Schulze}(2020)}]{Martini:2019lsi}
\bibinfo{author}{\bibfnamefont{T.}~\bibnamefont{Martini}} \bibnamefont{and}
  \bibinfo{author}{\bibfnamefont{M.}~\bibnamefont{Schulze}},
  \bibinfo{journal}{JHEP} \textbf{\bibinfo{volume}{04}}, \bibinfo{pages}{017}
  (\bibinfo{year}{2020}), \eprint{1911.11244}.

\bibitem[{\citenamefont{Richard}(2014)}]{Richard:2014upa}
\bibinfo{author}{\bibfnamefont{F.}~\bibnamefont{Richard}}
  (\bibinfo{year}{2014}), \eprint{1403.2893}.

\bibitem[{\citenamefont{Caola and Melnikov}(2013)}]{Caola:2013yja}
\bibinfo{author}{\bibfnamefont{F.}~\bibnamefont{Caola}} \bibnamefont{and}
  \bibinfo{author}{\bibfnamefont{K.}~\bibnamefont{Melnikov}},
  \bibinfo{journal}{Phys. Rev.} \textbf{\bibinfo{volume}{D88}},
  \bibinfo{pages}{054024} (\bibinfo{year}{2013}), \eprint{1307.4935}.

\bibitem[{\citenamefont{Chen et~al.}(2014)\citenamefont{Chen, Cheng, Gainer,
  Korytov, Matchev, Milenovic, Mitselmakher, Park, Rinkevicius, and
  Snowball}}]{Chen:2013waa}
\bibinfo{author}{\bibfnamefont{M.}~\bibnamefont{Chen}},
  \bibinfo{author}{\bibfnamefont{T.}~\bibnamefont{Cheng}},
  \bibinfo{author}{\bibfnamefont{J.~S.} \bibnamefont{Gainer}},
  \bibinfo{author}{\bibfnamefont{A.}~\bibnamefont{Korytov}},
  \bibinfo{author}{\bibfnamefont{K.~T.} \bibnamefont{Matchev}},
  \bibinfo{author}{\bibfnamefont{P.}~\bibnamefont{Milenovic}},
  \bibinfo{author}{\bibfnamefont{G.}~\bibnamefont{Mitselmakher}},
  \bibinfo{author}{\bibfnamefont{M.}~\bibnamefont{Park}},
  \bibinfo{author}{\bibfnamefont{A.}~\bibnamefont{Rinkevicius}},
  \bibnamefont{and} \bibinfo{author}{\bibfnamefont{M.}~\bibnamefont{Snowball}},
  \bibinfo{journal}{Phys. Rev.} \textbf{\bibinfo{volume}{D89}},
  \bibinfo{pages}{034002} (\bibinfo{year}{2014}), \eprint{1310.1397}.

\bibitem[{\citenamefont{Campbell et~al.}(2014)\citenamefont{Campbell, Ellis,
  and Williams}}]{Campbell:2013una}
\bibinfo{author}{\bibfnamefont{J.~M.} \bibnamefont{Campbell}},
  \bibinfo{author}{\bibfnamefont{R.~K.} \bibnamefont{Ellis}}, \bibnamefont{and}
  \bibinfo{author}{\bibfnamefont{C.}~\bibnamefont{Williams}},
  \bibinfo{journal}{JHEP} \textbf{\bibinfo{volume}{04}}, \bibinfo{pages}{060}
  (\bibinfo{year}{2014}), \eprint{1311.3589}.

\bibitem[{\citenamefont{Coleppa et~al.}(2014)\citenamefont{Coleppa, Mandal, and
  Mitra}}]{Coleppa:2014qja}
\bibinfo{author}{\bibfnamefont{B.}~\bibnamefont{Coleppa}},
  \bibinfo{author}{\bibfnamefont{T.}~\bibnamefont{Mandal}}, \bibnamefont{and}
  \bibinfo{author}{\bibfnamefont{S.}~\bibnamefont{Mitra}},
  \bibinfo{journal}{Phys. Rev.} \textbf{\bibinfo{volume}{D90}},
  \bibinfo{pages}{055019} (\bibinfo{year}{2014}), \eprint{1401.4039}.

\bibitem[{\citenamefont{Gainer et~al.}(2015)\citenamefont{Gainer, Lykken,
  Matchev, Mrenna, and Park}}]{Gainer:2014hha}
\bibinfo{author}{\bibfnamefont{J.~S.} \bibnamefont{Gainer}},
  \bibinfo{author}{\bibfnamefont{J.}~\bibnamefont{Lykken}},
  \bibinfo{author}{\bibfnamefont{K.~T.} \bibnamefont{Matchev}},
  \bibinfo{author}{\bibfnamefont{S.}~\bibnamefont{Mrenna}}, \bibnamefont{and}
  \bibinfo{author}{\bibfnamefont{M.}~\bibnamefont{Park}},
  \bibinfo{journal}{Phys. Rev.} \textbf{\bibinfo{volume}{D91}},
  \bibinfo{pages}{035011} (\bibinfo{year}{2015}), \eprint{1403.4951}.

\bibitem[{\citenamefont{Azatov et~al.}(2015)\citenamefont{Azatov, Grojean,
  Paul, and Salvioni}}]{Azatov:2014jga}
\bibinfo{author}{\bibfnamefont{A.}~\bibnamefont{Azatov}},
  \bibinfo{author}{\bibfnamefont{C.}~\bibnamefont{Grojean}},
  \bibinfo{author}{\bibfnamefont{A.}~\bibnamefont{Paul}}, \bibnamefont{and}
  \bibinfo{author}{\bibfnamefont{E.}~\bibnamefont{Salvioni}},
  \bibinfo{journal}{Zh. Eksp. Teor. Fiz.} \textbf{\bibinfo{volume}{147}},
  \bibinfo{pages}{410} (\bibinfo{year}{2015}), \bibinfo{note}{[J. Exp. Theor.
  Phys.120,354(2015)]}, \eprint{1406.6338}.

\bibitem[{\citenamefont{Englert
  et~al.}(2015{\natexlab{a}})\citenamefont{Englert, Soreq, and
  Spannowsky}}]{Englert:2014ffa}
\bibinfo{author}{\bibfnamefont{C.}~\bibnamefont{Englert}},
  \bibinfo{author}{\bibfnamefont{Y.}~\bibnamefont{Soreq}}, \bibnamefont{and}
  \bibinfo{author}{\bibfnamefont{M.}~\bibnamefont{Spannowsky}},
  \bibinfo{journal}{JHEP} \textbf{\bibinfo{volume}{05}}, \bibinfo{pages}{145}
  (\bibinfo{year}{2015}{\natexlab{a}}), \eprint{1410.5440}.

\bibitem[{\citenamefont{Li et~al.}(2015)\citenamefont{Li, Li, Shao, and
  Wang}}]{Li:2015jva}
\bibinfo{author}{\bibfnamefont{C.~S.} \bibnamefont{Li}},
  \bibinfo{author}{\bibfnamefont{H.~T.} \bibnamefont{Li}},
  \bibinfo{author}{\bibfnamefont{D.~Y.} \bibnamefont{Shao}}, \bibnamefont{and}
  \bibinfo{author}{\bibfnamefont{J.}~\bibnamefont{Wang}},
  \bibinfo{journal}{JHEP} \textbf{\bibinfo{volume}{08}}, \bibinfo{pages}{065}
  (\bibinfo{year}{2015}), \eprint{1504.02388}.

\bibitem[{\citenamefont{Englert
  et~al.}(2015{\natexlab{b}})\citenamefont{Englert, Low, and
  Spannowsky}}]{Englert:2015zra}
\bibinfo{author}{\bibfnamefont{C.}~\bibnamefont{Englert}},
  \bibinfo{author}{\bibfnamefont{I.}~\bibnamefont{Low}}, \bibnamefont{and}
  \bibinfo{author}{\bibfnamefont{M.}~\bibnamefont{Spannowsky}},
  \bibinfo{journal}{Phys. Rev.} \textbf{\bibinfo{volume}{D91}},
  \bibinfo{pages}{074029} (\bibinfo{year}{2015}{\natexlab{b}}),
  \eprint{1502.04678}.

\bibitem[{\citenamefont{Azatov et~al.}(2016)\citenamefont{Azatov, Grojean,
  Paul, and Salvioni}}]{Azatov:2016xik}
\bibinfo{author}{\bibfnamefont{A.}~\bibnamefont{Azatov}},
  \bibinfo{author}{\bibfnamefont{C.}~\bibnamefont{Grojean}},
  \bibinfo{author}{\bibfnamefont{A.}~\bibnamefont{Paul}}, \bibnamefont{and}
  \bibinfo{author}{\bibfnamefont{E.}~\bibnamefont{Salvioni}},
  \bibinfo{journal}{JHEP} \textbf{\bibinfo{volume}{09}}, \bibinfo{pages}{123}
  (\bibinfo{year}{2016}), \eprint{1608.00977}.

\bibitem[{\citenamefont{Goncalves
  et~al.}(2018{\natexlab{a}})\citenamefont{Goncalves, Han, and
  Mukhopadhyay}}]{Goncalves:2017iub}
\bibinfo{author}{\bibfnamefont{D.}~\bibnamefont{Goncalves}},
  \bibinfo{author}{\bibfnamefont{T.}~\bibnamefont{Han}}, \bibnamefont{and}
  \bibinfo{author}{\bibfnamefont{S.}~\bibnamefont{Mukhopadhyay}},
  \bibinfo{journal}{Phys. Rev. Lett.} \textbf{\bibinfo{volume}{120}},
  \bibinfo{pages}{111801} (\bibinfo{year}{2018}{\natexlab{a}}),
  \bibinfo{note}{[Erratum: Phys. Rev. Lett.121,no.7,079902(2018)]},
  \eprint{1710.02149}.

\bibitem[{\citenamefont{Lee et~al.}(2019{\natexlab{a}})\citenamefont{Lee, Park,
  and Qian}}]{Lee:2018fxj}
\bibinfo{author}{\bibfnamefont{S.~J.} \bibnamefont{Lee}},
  \bibinfo{author}{\bibfnamefont{M.}~\bibnamefont{Park}}, \bibnamefont{and}
  \bibinfo{author}{\bibfnamefont{Z.}~\bibnamefont{Qian}},
  \bibinfo{journal}{Phys. Rev.} \textbf{\bibinfo{volume}{D100}},
  \bibinfo{pages}{011702} (\bibinfo{year}{2019}{\natexlab{a}}),
  \eprint{1812.02679}.

\bibitem[{\citenamefont{Goncalves
  et~al.}(2018{\natexlab{b}})\citenamefont{Goncalves, Han, and
  Mukhopadhyay}}]{Goncalves:2018pkt}
\bibinfo{author}{\bibfnamefont{D.}~\bibnamefont{Goncalves}},
  \bibinfo{author}{\bibfnamefont{T.}~\bibnamefont{Han}}, \bibnamefont{and}
  \bibinfo{author}{\bibfnamefont{S.}~\bibnamefont{Mukhopadhyay}},
  \bibinfo{journal}{Phys. Rev.} \textbf{\bibinfo{volume}{D98}},
  \bibinfo{pages}{015023} (\bibinfo{year}{2018}{\natexlab{b}}),
  \eprint{1803.09751}.

\bibitem[{\citenamefont{He et~al.}(2019)\citenamefont{He, Wan, and
  Wang}}]{He:2019kgh}
\bibinfo{author}{\bibfnamefont{H.-R.} \bibnamefont{He}},
  \bibinfo{author}{\bibfnamefont{X.}~\bibnamefont{Wan}}, \bibnamefont{and}
  \bibinfo{author}{\bibfnamefont{Y.-K.} \bibnamefont{Wang}}
  (\bibinfo{year}{2019}), \eprint{1902.04756}.

\bibitem[{\citenamefont{Glover and van~der Bij}(1989)}]{Glover:1988rg}
\bibinfo{author}{\bibfnamefont{E.~W.~N.} \bibnamefont{Glover}}
  \bibnamefont{and} \bibinfo{author}{\bibfnamefont{J.~J.} \bibnamefont{van~der
  Bij}}, \bibinfo{journal}{Nucl. Phys.} \textbf{\bibinfo{volume}{B321}},
  \bibinfo{pages}{561} (\bibinfo{year}{1989}).

\bibitem[{\citenamefont{Hahn}(2001)}]{Hahn:2000kx}
\bibinfo{author}{\bibfnamefont{T.}~\bibnamefont{Hahn}},
  \bibinfo{journal}{Comput. Phys. Commun.} \textbf{\bibinfo{volume}{140}},
  \bibinfo{pages}{418} (\bibinfo{year}{2001}), \eprint{hep-ph/0012260}.

\bibitem[{\citenamefont{Shtabovenko et~al.}(2016)\citenamefont{Shtabovenko,
  Mertig, and Orellana}}]{Shtabovenko:2016sxi}
\bibinfo{author}{\bibfnamefont{V.}~\bibnamefont{Shtabovenko}},
  \bibinfo{author}{\bibfnamefont{R.}~\bibnamefont{Mertig}}, \bibnamefont{and}
  \bibinfo{author}{\bibfnamefont{F.}~\bibnamefont{Orellana}},
  \bibinfo{journal}{Comput. Phys. Commun.} \textbf{\bibinfo{volume}{207}},
  \bibinfo{pages}{432} (\bibinfo{year}{2016}), \eprint{1601.01167}.

\bibitem[{\citenamefont{Alwall et~al.}(2014)\citenamefont{Alwall, Frederix,
  Frixione, Hirschi, Maltoni, Mattelaer, Shao, Stelzer, Torrielli, and
  Zaro}}]{Alwall:2014hca}
\bibinfo{author}{\bibfnamefont{J.}~\bibnamefont{Alwall}},
  \bibinfo{author}{\bibfnamefont{R.}~\bibnamefont{Frederix}},
  \bibinfo{author}{\bibfnamefont{S.}~\bibnamefont{Frixione}},
  \bibinfo{author}{\bibfnamefont{V.}~\bibnamefont{Hirschi}},
  \bibinfo{author}{\bibfnamefont{F.}~\bibnamefont{Maltoni}},
  \bibinfo{author}{\bibfnamefont{O.}~\bibnamefont{Mattelaer}},
  \bibinfo{author}{\bibfnamefont{H.~S.} \bibnamefont{Shao}},
  \bibinfo{author}{\bibfnamefont{T.}~\bibnamefont{Stelzer}},
  \bibinfo{author}{\bibfnamefont{P.}~\bibnamefont{Torrielli}},
  \bibnamefont{and} \bibinfo{author}{\bibfnamefont{M.}~\bibnamefont{Zaro}},
  \bibinfo{journal}{JHEP} \textbf{\bibinfo{volume}{07}}, \bibinfo{pages}{079}
  (\bibinfo{year}{2014}), \eprint{1405.0301}.

\bibitem[{\citenamefont{Aaboud et~al.}(2018)}]{Aaboud:2018puo}
\bibinfo{author}{\bibfnamefont{M.}~\bibnamefont{Aaboud}} \bibnamefont{et~al.}
  (\bibinfo{collaboration}{ATLAS}), \bibinfo{journal}{Phys. Lett.}
  \textbf{\bibinfo{volume}{B786}}, \bibinfo{pages}{223} (\bibinfo{year}{2018}),
  \eprint{1808.01191}.

\bibitem[{\citenamefont{Sjostrand et~al.}(2008)\citenamefont{Sjostrand, Mrenna,
  and Skands}}]{Sjostrand:2007gs}
\bibinfo{author}{\bibfnamefont{T.}~\bibnamefont{Sjostrand}},
  \bibinfo{author}{\bibfnamefont{S.}~\bibnamefont{Mrenna}}, \bibnamefont{and}
  \bibinfo{author}{\bibfnamefont{P.~Z.} \bibnamefont{Skands}},
  \bibinfo{journal}{Comput. Phys. Commun.} \textbf{\bibinfo{volume}{178}},
  \bibinfo{pages}{852} (\bibinfo{year}{2008}), \eprint{0710.3820}.

\bibitem[{\citenamefont{de~Favereau et~al.}(2014)\citenamefont{de~Favereau,
  Delaere, Demin, Giammanco, Lemaitre, Mertens, and
  Selvaggi}}]{deFavereau:2013fsa}
\bibinfo{author}{\bibfnamefont{J.}~\bibnamefont{de~Favereau}},
  \bibinfo{author}{\bibfnamefont{C.}~\bibnamefont{Delaere}},
  \bibinfo{author}{\bibfnamefont{P.}~\bibnamefont{Demin}},
  \bibinfo{author}{\bibfnamefont{A.}~\bibnamefont{Giammanco}},
  \bibinfo{author}{\bibfnamefont{V.}~\bibnamefont{Lemaitre}},
  \bibinfo{author}{\bibfnamefont{A.}~\bibnamefont{Mertens}}, \bibnamefont{and}
  \bibinfo{author}{\bibfnamefont{M.}~\bibnamefont{Selvaggi}}
  (\bibinfo{collaboration}{DELPHES 3}), \bibinfo{journal}{JHEP}
  \textbf{\bibinfo{volume}{02}}, \bibinfo{pages}{057} (\bibinfo{year}{2014}),
  \eprint{1307.6346}.

\bibitem[{\citenamefont{Caola et~al.}(2015)\citenamefont{Caola, Melnikov,
  Rontsch, and Tancredi}}]{Caola:2015psa}
\bibinfo{author}{\bibfnamefont{F.}~\bibnamefont{Caola}},
  \bibinfo{author}{\bibfnamefont{K.}~\bibnamefont{Melnikov}},
  \bibinfo{author}{\bibfnamefont{R.}~\bibnamefont{Rontsch}}, \bibnamefont{and}
  \bibinfo{author}{\bibfnamefont{L.}~\bibnamefont{Tancredi}},
  \bibinfo{journal}{Phys. Rev.} \textbf{\bibinfo{volume}{D92}},
  \bibinfo{pages}{094028} (\bibinfo{year}{2015}), \eprint{1509.06734}.

\bibitem[{\citenamefont{Cascioli et~al.}(2014)\citenamefont{Cascioli, Gehrmann,
  Grazzini, Kallweit, Maierhofer, von Manteuffel, Pozzorini, Rathlev, Tancredi,
  and Weihs}}]{Cascioli:2014yka}
\bibinfo{author}{\bibfnamefont{F.}~\bibnamefont{Cascioli}},
  \bibinfo{author}{\bibfnamefont{T.}~\bibnamefont{Gehrmann}},
  \bibinfo{author}{\bibfnamefont{M.}~\bibnamefont{Grazzini}},
  \bibinfo{author}{\bibfnamefont{S.}~\bibnamefont{Kallweit}},
  \bibinfo{author}{\bibfnamefont{P.}~\bibnamefont{Maierhofer}},
  \bibinfo{author}{\bibfnamefont{A.}~\bibnamefont{von Manteuffel}},
  \bibinfo{author}{\bibfnamefont{S.}~\bibnamefont{Pozzorini}},
  \bibinfo{author}{\bibfnamefont{D.}~\bibnamefont{Rathlev}},
  \bibinfo{author}{\bibfnamefont{L.}~\bibnamefont{Tancredi}}, \bibnamefont{and}
  \bibinfo{author}{\bibfnamefont{E.}~\bibnamefont{Weihs}},
  \bibinfo{journal}{Phys. Lett.} \textbf{\bibinfo{volume}{B735}},
  \bibinfo{pages}{311} (\bibinfo{year}{2014}), \eprint{1405.2219}.

\bibitem[{\citenamefont{Heinrich et~al.}(2018)\citenamefont{Heinrich, Jahn,
  Jones, Kerner, and Pires}}]{Heinrich:2017bvg}
\bibinfo{author}{\bibfnamefont{G.}~\bibnamefont{Heinrich}},
  \bibinfo{author}{\bibfnamefont{S.}~\bibnamefont{Jahn}},
  \bibinfo{author}{\bibfnamefont{S.~P.} \bibnamefont{Jones}},
  \bibinfo{author}{\bibfnamefont{M.}~\bibnamefont{Kerner}}, \bibnamefont{and}
  \bibinfo{author}{\bibfnamefont{J.}~\bibnamefont{Pires}},
  \bibinfo{journal}{JHEP} \textbf{\bibinfo{volume}{03}}, \bibinfo{pages}{142}
  (\bibinfo{year}{2018}), \eprint{1710.06294}.

\bibitem[{\citenamefont{Grazzini et~al.}(2019)\citenamefont{Grazzini, Kallweit,
  Wiesemann, and Yook}}]{Grazzini:2018owa}
\bibinfo{author}{\bibfnamefont{M.}~\bibnamefont{Grazzini}},
  \bibinfo{author}{\bibfnamefont{S.}~\bibnamefont{Kallweit}},
  \bibinfo{author}{\bibfnamefont{M.}~\bibnamefont{Wiesemann}},
  \bibnamefont{and} \bibinfo{author}{\bibfnamefont{J.~Y.} \bibnamefont{Yook}},
  \bibinfo{journal}{JHEP} \textbf{\bibinfo{volume}{03}}, \bibinfo{pages}{070}
  (\bibinfo{year}{2019}), \eprint{1811.09593}.

\bibitem[{\citenamefont{Kallweit and Wiesemann}(2018)}]{Kallweit:2018nyv}
\bibinfo{author}{\bibfnamefont{S.}~\bibnamefont{Kallweit}} \bibnamefont{and}
  \bibinfo{author}{\bibfnamefont{M.}~\bibnamefont{Wiesemann}},
  \bibinfo{journal}{Phys. Lett.} \textbf{\bibinfo{volume}{B786}},
  \bibinfo{pages}{382} (\bibinfo{year}{2018}), \eprint{1806.05941}.

\bibitem[{\citenamefont{Agarwal and Von~Manteuffel}(2019)}]{Agarwal:2019rag}
\bibinfo{author}{\bibfnamefont{B.}~\bibnamefont{Agarwal}} \bibnamefont{and}
  \bibinfo{author}{\bibfnamefont{A.}~\bibnamefont{Von~Manteuffel}}, in
  \emph{\bibinfo{booktitle}{{14th International Symposium on Radiative
  Corrections: Application of Quantum Field Theory to Phenomenology (RADCOR
  2019) Avignon, France, September 8-13, 2019}}} (\bibinfo{year}{2019}),
  \eprint{1912.08794}.

\bibitem[{\citenamefont{Cowan et~al.}(2011)\citenamefont{Cowan, Cranmer, Gross,
  and Vitells}}]{Cowan:2010js}
\bibinfo{author}{\bibfnamefont{G.}~\bibnamefont{Cowan}},
  \bibinfo{author}{\bibfnamefont{K.}~\bibnamefont{Cranmer}},
  \bibinfo{author}{\bibfnamefont{E.}~\bibnamefont{Gross}}, \bibnamefont{and}
  \bibinfo{author}{\bibfnamefont{O.}~\bibnamefont{Vitells}},
  \bibinfo{journal}{Eur. Phys. J.} \textbf{\bibinfo{volume}{C71}},
  \bibinfo{pages}{1554} (\bibinfo{year}{2011}), \bibinfo{note}{[Erratum: Eur.
  Phys. J.C73,2501(2013)]}, \eprint{1007.1727}.

\bibitem[{\citenamefont{Aad et~al.}(2020)}]{Aad:2020wog}
\bibinfo{author}{\bibfnamefont{G.}~\bibnamefont{Aad}} \bibnamefont{et~al.}
  (\bibinfo{collaboration}{ATLAS}) (\bibinfo{year}{2020}), \eprint{2002.07546}.

\bibitem[{\citenamefont{Sirunyan et~al.}(2019)}]{Sirunyan:2018zgs}
\bibinfo{author}{\bibfnamefont{A.~M.} \bibnamefont{Sirunyan}}
  \bibnamefont{et~al.} (\bibinfo{collaboration}{CMS}), \bibinfo{journal}{Phys.
  Rev. Lett.} \textbf{\bibinfo{volume}{122}}, \bibinfo{pages}{132003}
  (\bibinfo{year}{2019}), \eprint{1812.05900}.

\bibitem[{\citenamefont{Sirunyan et~al.}(2020)}]{CMS:2019too}
\bibinfo{author}{\bibfnamefont{A.~M.} \bibnamefont{Sirunyan}}
  \bibnamefont{et~al.} (\bibinfo{collaboration}{CMS}), \bibinfo{journal}{JHEP}
  \textbf{\bibinfo{volume}{03}}, \bibinfo{pages}{056} (\bibinfo{year}{2020}),
  \eprint{1907.11270}.

\bibitem[{\citenamefont{Aaboud et~al.}(2019)}]{Aaboud:2019njj}
\bibinfo{author}{\bibfnamefont{M.}~\bibnamefont{Aaboud}} \bibnamefont{et~al.}
  (\bibinfo{collaboration}{ATLAS}), \bibinfo{journal}{Phys. Rev.}
  \textbf{\bibinfo{volume}{D99}}, \bibinfo{pages}{072009}
  (\bibinfo{year}{2019}), \eprint{1901.03584}.

\bibitem[{\citenamefont{Mangano et~al.}(2017)}]{Mangano:2016jyj}
\bibinfo{author}{\bibfnamefont{M.}~\bibnamefont{Mangano}} \bibnamefont{et~al.},
  \bibinfo{journal}{CERN Yellow Rep.} pp. \bibinfo{pages}{1--254}
  (\bibinfo{year}{2017}), \eprint{1607.01831}.

\bibitem[{\citenamefont{Vos}(2017)}]{Vos:2017jxz}
\bibinfo{author}{\bibfnamefont{M.}~\bibnamefont{Vos}}, \bibinfo{journal}{PoS}
  \textbf{\bibinfo{volume}{EPS-HEP2017}}, \bibinfo{pages}{471}
  (\bibinfo{year}{2017}).

\bibitem[{\citenamefont{Lee et~al.}(2019{\natexlab{b}})\citenamefont{Lee,
  Chanon, Levin, Li, Lu, Li, and Mao}}]{Lee:2019nhm}
\bibinfo{author}{\bibfnamefont{J.}~\bibnamefont{Lee}},
  \bibinfo{author}{\bibfnamefont{N.}~\bibnamefont{Chanon}},
  \bibinfo{author}{\bibfnamefont{A.}~\bibnamefont{Levin}},
  \bibinfo{author}{\bibfnamefont{J.}~\bibnamefont{Li}},
  \bibinfo{author}{\bibfnamefont{M.}~\bibnamefont{Lu}},
  \bibinfo{author}{\bibfnamefont{Q.}~\bibnamefont{Li}}, \bibnamefont{and}
  \bibinfo{author}{\bibfnamefont{Y.}~\bibnamefont{Mao}},
  \bibinfo{journal}{Phys. Rev.} \textbf{\bibinfo{volume}{D100}},
  \bibinfo{pages}{116010} (\bibinfo{year}{2019}{\natexlab{b}}),
  \eprint{1908.05196}.

\end{thebibliography}

\end{document}